# Rapid Simultaneous Mapping of Total and Myelin Water Content, $T_1$ and $T_2^*$ in Multiple Sclerosis

V.Arhelger[1], F. Kasten[1], D. Gliedstein[2], M.S.Lafontaine[1], V.Tonkova[1], D. Holz[1,3], A. Böer[4], J. Schenk[2], and H. Neeb[1,3,*]

[1] University of Applied Sciences Koblenz, RheinAhrCampus Remagen, 53424 Remagen / Germany.
[2] Radiologisches Institut Hohenzollernstrasse, 56068 Koblenz / Germany.
[3] Institute for Medical Engineering and Information Processing – MTI Mittelrhein, University of Koblenz, 56070 Koblenz / Germany.
[4] Neurologie Dr. Böer, 56068 Koblenz / Germany
*Address correspondence to: Professor Dr. H. Neeb, RheinAhrCampus Remagen, University of Applied Sciences Koblenz, 53424 Remagen, Germany, Tel: +49 2642 932 443, Email: neeb@rheinahrcampus.de

## ABSTRACT

It is long discussed that quantitative magnetic resonance imaging might provide a more specific insight into disease process, progression and therapeutic response of multiple sclerosis. We present an extension of a previously published approach for the simultaneous mapping of brain $T_1$, $T_2^*$ and total water content. In addition to those three parameters, the method presented in the current work allows for the measurement of myelin bound water content, a surrogate marker of tissue myelination. Myelin water was measured based on its distinct relaxation with reduced $T_2^*$, resulting in a multiexponential decay signal. However, only 10 points could be acquired on the relaxation curve within a maximum echo time of <40ms as the quantitative protocol has been adapted previously for fast acquisitions with whole brain coverage. The sparse sampling required an adaption of the optimisation approach with additional constraints necessary in order to obtain reliable results. Therefore, the corresponding pool fractions were determined using linear optimisation instead of the standard nonnegative least squares (NNLS). The whole approach including the proper choice of constraints was optimised and validated in simulation studies. Furthermore, it was shown that total water content measurement based on a single exponential model is significantly biased in tissue which shows biexponential behaviour such as white matter. Therefore, the corresponding reconstruction algorithm was refined in the current work in order to reduce the systematic measurement error. Furthermore, the independent determination of total water content allows for an absolute quantification of myelin water content, resulting in a more reliable measurement in oedemateous tissue. Using the approach developed, whole brain maps of $T_1$, $T_2^*$, total and myelin water content were acquired in 12 patients suffering from multiple sclerosis with mild disease grade. The results obtained were consistent with previous reports for the parameters investigated. This demonstrates that simultaneous whole brain mapping of $T_1$, $T_2^*$, total and myelin water content is feasible on almost any modern MR scanner. With an acquisition time of 10 minutes only, the presented multidimensional quantitative MRI protocol provides an interesting option for the clinical assessment and monitoring of multiple sclerosis, even on a routine basis.

## INTRODUCTION

Quantitative MRI of multiple sclerosis has gained increased attention over the past few years. This is in great parts due to the availability of new methods that allow for the measurement of parameters such as relaxation times or diffusion anisotropy with high precision and accuracy so that even small pathological changes can be detected (Bakshi et al., 2008; Neema et al., 2007a; Neema et al., 2007b).

One of the most promising parameters investigated refers to water content trapped by the myelin sheet, especially in white matter. Much of the pioneering work has been published by the Vancouver group who very convincingly demonstrated that quantitative MRI of myelin is feasible (Laule et al., 2004; Laule et al., 2006; Laule et al., 2007; Laule et al., 2008; Kolind et al., 2009; Meyer et al., 2009; Minty et al., 2009; Vavasour et al., 2009). Their method is based on a multiexponential analysis of the $T_2$ decay curve in order to quantify myelin-trapped water by its distinct, i.e. short, relaxation time as compared to other compartments. Myelin water content is defined by the amplitude ratio of the short $T_2$ pool with respect to the total amplitude of all pools. As this approach is based on the calculation of ratios, a decreased myelin water content ratio is compatible with an increased numerator or a decreased enumerator. Therefore, the method provides more indirect information about demyelination processes in multiple sclerosis as the cofounding factor of an increased total water content is not determined independently. However, it is not to be unexpected that total water content increases in white matter of MS patients as a $T_1$ is prolongation in this group was independently confirmed (Vrenken et al., 2006a; Vrenken et al., 2006b; Manfredonia et al., 2007). As $T_1$ correlates with total proton density, water content might be increased accordingly. This would result in an overestimation of demyelination effects in regions with simultaneous presence of significant oedema.

As previously shown, the average total water content in white matter of healthy subjects ranges between 68% and 72% without significant age or gender effects present (Neeb et al, 2006b). However, significant regional variations were observed. If these variations are, at least in parts, dominated by changes in the free water pool, the corresponding myelin water content measurement will be biased accordingly. Using the brain water model discussed by Laule et al. (2004), it was shown by the authors that a 4% change of water content could explain the observed myelin water content reduction in the MS subjects studied. However, that is exactly the same order of magnitude for physiological variations of white matter water content in a healthy population. Therefore, an absolute quantification of myelin water content could provide a further enhancement of sensitivity so that even smaller changes might be detectable. Vavasour et al. (2009) have recently published results for the simultaneous measurement of total and myelin water content in selected MS lesions. However, their method did not correct for receiver coil and $B_1$ inhomogeneities which is vital in order to minimise systematic errors in absolute water content mapping (Tofts, 2003; Neeb et al., 2006a; Neeb et al, 2008; Warntjes et al. 2008). Furthermore, the protocol presented allows only for a single slice measurement within an acquisition time $T_{acq} \approx 25min$. Therefore, faster protocols with multi slice capability are required for a more widespread clinical application of quantitative myelin water content mapping. Consequently, several approaches with higher spatial coverage and reduced acquisition time have been published in the recent past (Oh et al., 2007; Du et al., 2007; Deoni et al., 2008; Hwang et al., 2010). Oh et al. (2007) have presented a method for myelin water content mapping based on a fast $T_2$-prep spiral sequence with 12 points acquired on the $T_2$ relaxation curve and maximum echo time of $294\,ms$. They were able to acquire 16 slices with a voxel volume of $2x2x5mm^3$ within 10 minutes. The sampling density of the relaxation curve was increased to 126 points by Du et al. (2007) who demonstrated the feasibility to measure myelin water content employing a 3-pool model and $T_2^*$ decay data. However, only 5 slices with a voxel volume of $0.78x0.78x5mm^3$ were acquired in 8.7 minutes. The corresponding protocol has recently been adapted for *in vivo* imaging with a slightly increased measurement time per slice (Hwang et al, 2010). Even though multiple slices can be acquired using those approaches, the acquisition time of approx. 40-60s/slice still prevents a measurement with full brain coverage and reasonable slice thickness within clinically relevant measurement times. This is in contrast to the 3D approach described by Deoni et al. (2008) where 60 slices can be acquired in 16 minutes. However, the myelin water fraction reported in this study is significantly higher in both white and grey matter than the corresponding results obtained by other groups (Laule et al., 2004; Oh et al., 2007; Du et al., 2007; Kolind et al., 2009; Vavasour et al., 2009; Hwang et al., 2010). Even though multiple parameters including the proton residence time can be determined using this elegant approach, more work is required to investigate the origin of the systematic shift before the method can be applied for an accurate assessment of tissue myelin water content.

None of the fast multislice methods provides a combined acquisition of myelin and absolute water content in order to better distinguish the relative contribution of demyelination and oedema in pathological brain tissue. Therefore, we have investigated the feasibility of the approach published by Neeb et al. (2008) for a simultaneous mapping of absolute and myelin water content. The protocol is based on the acquisition of a multi echo gradient echo (MEGE) sequence and allows for the mapping of tissue specific $T_1$, $T_2^*$ and total water content with full brain coverage and $1x1x2mm^3$ voxel size. The MEGE protocol is used to sample 10 points on the $T_2^*$ relaxation curve with a maximum echo time of approx. $40ms$. As, however, the majority of attempts to measure myelin water content are based on relaxation data with >10 points and echo times up to $320ms$ (Laule et al., 2004; Oh et al., 2007; Du et al., 2007; Hwang et al., 2010), it is not directly obvious that myelin water mapping is also feasible using the protocol given in (Neeb et al., 2008). Most importantly, many investigators have employed a regularized nonnegative least squares optimisation (NNLS) in order to estimate the relative pool fractions. Due to the sparse sampling of the relaxation curve, this approach was not feasible for the echo times regime measured in the current study. Therefore, a linear optimisation (LP) algorithm had to be employed where additional constraints can easily be incorporated. However, each additional assumption is a potential source for systematic error so that LP constraints were systematically changed in order to evaluate and minimise any possible bias. The optimisation approach was extensively validated in simulation studies employing a two pool relaxation model (myelin and “rest”). Given the proper choice of model parameters, the results demonstrate that a precise and accurate estimation of the corresponding pool fractions is feasible even with an extremely sparse sampling of the relaxation curve.

Moreover, the total water content measurement can be refined based on the pool fractions obtained by the linear optimisation approach. In tissue with significant biexponential $T_2^*$ decay, a single exponential fit might compromise the accurate extrapolation of the decay curve to $TE = 0ms$, which is required for water content mapping. The bias introduced by the single exponential model was studied and a refined extrapolation algorithm for absolute water content mapping has been developed, resulting in a more accurate measurement, especially in white matter. The whole approach was applied in a group of 12 patients suffering from multiple sclerosis in order to demonstrate the feasibility of simultaneous mapping of total and myelin water content, $T_1$ and $T_2^*$ *in vivo* with full brain coverage in less than 10 minutes on a clinical MR system.

## METHODS

### *In vivo* Measurements

Simultaneous quantitative mapping of absolute water content, $T_1$ and $T_2^*$ was performed in a group of 12 patients suffering from multiple sclerosis as described in (Neeb et al, 2008). The group consists of 6 male and 6 female subjects with an average age of 36.2 and 41 years, respectively. Disease grade for each individual subject was evaluated using the expanded disability status state (EDSS) and two groups with EDSS=0 (N=6) and EDSS=1 (N=6) were investigated. The quantitative MR protocol was based on the acquisition of two multi echo gradient echo sequences (MEGE) and tree echo-planar imaging scans. In the current work, however, the in-house developed sequence termed QUTE (Neeb et al, 2008) was replaced by an identical sequence ("gre") provided by the scanner manufacturer. The remaining part of the approach, i.e. sequence and timing parameters, was left unchanged. The changed protocol based on the "gre" sequence has been extensively validated in phantom studies at 3T (Tonkova, 2009). As previously shown, quantitative maps with high accuracy and precision can be obtained in less than 10 minutes for 50 slices with $2mm$ thickness and $1mm^2$ in plane resolution. However, the main focus of the current work is to demonstrate that myelin water content can be determined in addition to the other parameters without changing the imaging protocol employed for data acquisition. Therefore, quantification of myelin water content was based on the measurement of 10 points on the $T_2^*$ decay curve using the same MEGE sequence as for the quantitative protocol. I.e., the echo time was 4.8ms with an echo spacing of 3.74ms, resulting in $TE_{max}$ of only $38.46ms$ (Neeb et al., 2008).

In addition to the quantitative protocol, the standard diagnostic procedure for MS outpatients in our institution consisting of dark-fluid, $T_2$-weighted and high resolution pre- and post-contrast $T_1$-weighted imaging was performed for each subject. However, the information acquired in this context has not been explored in the current work. All measurements were performed on a 3T TRIO system (Siemens Medical Solutions, Erlangen, Germany) equipped with an 8 channel head coil for signal reception.

**Simulations**

In order to evaluate accuracy and precision of the MEGE acquisition for myelin water content quantification, synthetic decay data were simulated. In the current work, a 2-pool model of white matter with two distinct relaxation regimes was assumed. This model has been extensively studied by Laule et al. (2006) who have successfully demonstrated that MR-based measurements of myelin water content agrees well with results obtained from histopathology. However, as our approach has to be optimized for *in vivo* measurements at 3T using gradient echo data, the range of relaxation times needs to be adapted with respect to values given in the literature. As those values are not known *a priori*, we most generally assume two distinct pools with relaxation times $T_2^{*,f}$ and $T_2^{*,s}$ and corresponding pool fractions of 10% and 90%, consistent with previous reports on myelin water content *in vivo* (Laule et al., 2004; Oh et al., 2007; Du et al., 2007; Kolind et al., 2009; Vavasour et al., 2009; Hwang et al., 2010). The resulting model for the decay signal is then given by

$$y_i^{meas} \equiv S(t_i) = \left(0.1 \cdot e^{-\frac{t_i}{T_2^{*,f}}} + 0.9 \cdot e^{-\frac{t_i}{T_2^{*,s}}}\right) + N\left(0, \frac{1}{SNR}\right) \quad i = 1, \ldots, 10\,, \qquad [1]$$

where $N\left(0, \frac{1}{SNR}\right)$ represents normally distributed random numbers with zero mean and standard deviation $1/SNR$. To be consistent with real measurements, the signal-to-noise ratio, SNR, was set to a value of 80 and 10 points with the same *TE* and echo spacing as the *in vivo* data were simulated. The identity $y_i^{meas} \equiv S(t_i)$ highlights the fact that simulated decay data are noisy and therefore correspond to data acquired in a real measurement.

**Linear Optimisation**

In the most general case, the MRI measurement can be expressed as data points $y_i^{meas}$ which are linear combinations of $n$ exponential functions, depending on echo time, $t_i$ and compartment specific relaxation time, $T_2^*(j)$

$$y_i^{meas} = \sum_{j=1}^{n} s_j \cdot e^{-t_i/T_2^*(j)}, i = 1, \cdots, 10. \quad [2]$$

Here, $s_j \equiv s(T_2^*(j)) \geq 0$ represents the relative fraction occupied by the $j$-th compartment with relaxation time $T_2^*(j)$. In general, $y_i^{meas}$ differs from the unknown, exact value $y_i^{real}$ by a factor $\Delta_i$ which is assumed to be normally distributed with zero mean and standard deviation $\sigma$. Therefore, $y_i^{meas}$ can be expressed as $y_i^{meas} = y_i^{real} + \Delta_i$, or, in a more compact vector notation,$y^{meas} := y^{real} + \Delta \in \mathbb{R}^m$. With

$$A := \begin{pmatrix} e^{-t_1/T_2(1)} & \cdots & e^{-t_1/T_2(n)} \\ \vdots & \ddots & \vdots \\ e^{-t_{10}/T_2(1)} & \cdots & e^{-t_{10}/T_2(n)} \end{pmatrix} \text{ and } s := \begin{pmatrix} s_1 \\ \vdots \\ s_n \end{pmatrix}, \quad [3]$$

Equation (2) can also be written as $A \cdot s = y^{meas}$. In order to estimate the relative volume fractions, our aim was to find a vector $s^*$ so that $||A \cdot s^* - y^{meas}||_2^2$ becomes minimal and $s^*$ can be regarded as a good approximation to $s^{real}$, where $s^{real}$ is given by $A \cdot s^{real} = y^{real}$. Because of the relatively sparse sampling of the $T_2^*$ relaxation curve, $s^*$ was determined based on a linear programming (LP) approach. This is due to the fact that linear programming offers the possibility to incorporate additional information about $s^{real}$ into the optimization problem. This typically results in better estimates of $s^{real}$ as compared to NNLS, which is, apart from the requirement $s_j \geq 0$, unconstrainted. However, every additional constraint is a potential source of systematic error so that a balance between small bias and precise estimate of $s$ has to be found.

As a two pool approach is followed in the current work, the amplitudes were constrained to two distinct relaxation regimes. More specifically, $s(T_2^*(j))$ has to obey

$$s\big(T_2^*(j)\big) < a_f \cdot rect\left(\frac{T_2^*(j) - \widehat{T_2^{*,f}}}{\widehat{w_f}}\right) + a_s \cdot rect\left(\frac{T_2^*(j) - \widehat{T_2^{*,s}}}{\widehat{w_s}}\right), \quad [4]$$

with $rect(x) = 1$ for $|x| < 0.5$ and zero elsewhere. Here, $\widehat{T_2^{*,f}}$ ($\widehat{T_2^{*,s}}$) determines the position and $\widehat{w_f}$ ($\widehat{w_s}$) the total width of the interval where a nonzero amplitude was allowed for the fast and slow relaxing regime, respectively. The corresponding parameters, $a_f$ and $a_s$, control the desired maximum amplitudes. Equation (4) cuts the whole $T_2^*$ search space into two distinct intervals. No optimisation was performed for $T_2^*(j)$ values outside those two intervals in order to speed up calculations. In addition to the upper amplitude bound given by Equation (4), nonnegative lower amplitudes, $s_j \geq 0$, were required for the whole range of $T_2^*$ optimised.

Equation (4) contains six parameters, $\left(\widehat{T_2^{*,f}}, \widehat{T_2^{*,s}}\, \widehat{w_f}, \widehat{w_s}, a_f, a_s\right)$, which are in general not *a priori* known. Therefore, the corresponding parameter vector has to be estimated based on prior information or assumptions made about the relaxation process, i.e. about the position and the relative amplitudes of the two relaxation regimes. However, part of the estimation might be data driven which means that information extracted from the measured $T_2^*$ decay curve can be incorporated into the definition of the free parameters in Equation (4). Here, it is important to note the dominating effect of the slow relaxation compartment on the measured decay curve due to the small fraction of the myelin pool which accounts for only approx. 10% of the whole FID amplitude. This implies, that a single exponential fit of the decay curve will in general yield a result which is close to $\widehat{T_2^{*,s}}$, the relaxation time of the slow relaxing pool. Therefore, in order to estimate $\widehat{T_2^{*,s}}$ from the data, the decay curve of each point was first fitted using a single exponential decay model. The fit results in the determination of the two parameters $T_{2,SE}^*$ and $S_{0,SE}$, the relaxation time and the signal intensity at zero echo time, respectively. The index *SE* is used to highlight the fact that a single exponential fit has been employed to extract those parameters. $T_{2,SE}^*$ is not an accurate estimate of $\widehat{T_2^{*,s}}$ as the effect of the fast relaxing pool is still contained in that parameter. However, the accuracy can be improved by simulating the effect of the fast relaxation time on $T_{2,SE}^*$. Therefore, biexponential decay curves were modelled according to Equation (1) for different values of $T_2^{*,s}$ in the range between $25ms$ and $85ms$. The fast relaxation time, $T_2^{*,f}$, was fixed to $15ms$ which is at the lower end of the typical range given in the literature (Laule et al., 2004; Laule et al., 2007b; Kolind et al., 2009). This is due to the fact that most of the previously reported measurements were performed at 1.5T using spin echo sequences. It may therefore be anticipated that the relaxation times measured at 3T based on $T_2^*$ decay are shorter than the corresponding values given in the literature. For each $T_2^{*,s}$, 10000 samples were independently drawn from Equation (1) and the average of $T_{2,SE}^*$ was determined. This provides a direct mapping between $T_{2,SE}^*$ and $T_2^{*,s}$ which can be formally expressed by a function $T_2^{*,s} = f(T_{2,SE}^*)$. The so obtained functional dependence was then employed to define the centre position of the slow relaxation interval in Equation (4), $\widehat{T_2^{*,s}}$, by first performing a single exponential fit of each decay curve. Using the resulting single exponential relaxation time, $T^*_{2,SE}$, the unknown $\widehat{T_2^{*,s}}$ can then readily be obtained using $\widehat{T_2^{*,s}} \equiv T_2^{*,s} = f(T^*_{2,SE})$.

Furthermore, $S_{0,SE}$ was employed to define the maximum amplitudes of the fast and slow relaxing compartments as $a_f = 0.5 \cdot S_{0,SE}$ and $a_s = 2 \cdot S_{0,SE}$, respectively. Those are weak conditions which do not significantly constrain the estimation of myelin water content which is supposed to be <=20% (Laule et al., 2004; Oh et al., 2007; Du et al., 2007; Kolind et al., 2009; Vavasour et al., 2009; Hwang et al., 2010). The remaining three parameters in Equation (4), $\widehat{T_2^{*,f}}$, $\widehat{w_f}$ and $\widehat{w_s}$, were optimised in order to minimise systematic shifts as described below.

Based on the obtained amplitude spectrum, $s(T_2^*(j))$, the relative myelin water content, $W_{My}^{rel}$, is defined by the ratio between myelin amplitude, $S_{My} = \sum_{T_2^*(j) \in I^f} s(T_2^*(j))$, and total amplitude, $S_{Tot} = \sum_{T_2^*(j) \in (I^f \cup I^s)} s(T_2^*(j))$. Here, $I^f$ and $I^s$ represent the intervals defined by the first and the second term on the right hand side of Equation (4), respectively. Based on the relative myelin water content, $W_{My}^{rel}$, and the absolute water content, $W_{MR}$, the corresponding absolute myelin water content is given by $W_{My}^{abs} = W_{My}^{rel} \cdot W_{MR}$.

**Protocol Optimisation and Systematic Errors**

To estimate the remaining three free parameters in Equation (4), $\left(\widehat{T_2^{*,f}}, \widehat{w_f}, \widehat{w_s}\right)$, simulations were performed with the goal to determine a parameter vector which results in the smallest overall systematic error. As the systematic shift might depend on relaxation times of the fast and slow pool, synthetic data were simulated according to Equation (1) with $T_2^{*,f} \in [10, 20]\ ms$ and $T_2^{*,s} \in [25, 85]\ ms$, respectively. In addition, the three unknown parameters were changed as $\widehat{T_2^{*,f}} \in [10, 20]\ ms$, $\widehat{w_f} \in [5, 7.5, 10]\ ms$, $\widehat{w_s} \in [2,\ 5, 10, 15]\ ms$. For each of the possible combinations, the average relative myelin water content was determined from 10000 independent samplings of the relaxation curve according to Equation (1). The result can formally be expressed as $W_{My}^{rel}\left(T_2^{*,s}, T_2^{*,f}, \widehat{T_2^{*,f}}, \widehat{w_f}, \widehat{w_s}\right)$. As $T_2^{*,f}$ and $T_2^{*,s}$ will in general depend on spatial position in a real data set, e.g. due to magnetic field inhomogeneities or microstructural differences between tissues, our goal was to find the parameter vector $\left(\widehat{T_2^{*,f}}, \widehat{w_f}, \widehat{w_s}\right)$ which minimises the deviation of $W_{My}^{rel}$ from its real value averaged over all possible $T_2^{*,f}$ and $T_2^{*,s}$. An exhaustive search was performed to define the corresponding parameter set which will be referred to as *optimised myelin set* in the following.

As no further assumptions can be made about the real values of fast and slow relaxation times for a single voxel, this remaining uncertainty represents the main source of systematic measurement error which will therefore be a function of $T_2^{*,f}$ and $T_2^{*,s}$. Moreover, the result might be biased by the presence of further relaxation pools which are not considered by the two pool model employed in the current work. Laule and co-workers (2007a) have recently reported about a third, long relaxation time in human brain tissue. Therefore, possible effects due to a third component with long $T_2^*$ on the determination of myelin water content were evaluated. Based on the results given in (Laule et al., 2007a) for normal appearing white matter in MS subjects, a third component with an amplitude of 0.05 and $T_2^{*,3rd-pool}$ =200 ms was added to the simulated data model given in Equation (1). As described above, the average myelin water content was again determined for the different combinations of $T_2^{*,f}$ and $T_2^{*,s}$. We have chosen $T_2^{*,3rd-pool}$ to be $0.5 \cdot T_2$, with $T_2 \approx 400ms$ as reported in (Laule et al., 2007a) in order to account for the faster relaxation process employing gradient echo data at 3T.

**Magnetic field inhomogeneity correction**

The observed decay in a gradient echo experiment significantly deviates from a pure exponential behaviour in regions where strong magnetic field inhomogeneities exists across a voxel. In order to partly compensate such modulations, an approach similar to that described by Hwang et al. (2010) was followed in the current work. Briefly, magnetic field offset maps, $\Delta B(x, y, z)$, were created from a linear fit of the phase data acquired with the MEGE sequence which was also employed for myelin water mapping. The phase data were corrected for $2\pi$-fold effects prior fitting. Assuming a linear variation of the magnetic field across the voxel, a gradient map, $\Delta G_z(x, y, z)$, along the slice select direction was determined based on $\Delta B(x, y, z)$. As the voxel size in $z$-direction, $\Delta z$, is a factor two larger than in $x$ or $y$, the other two directions were neglected. The measured signal intensity at echo time $TE_i$ was then corrected by a multiplication of the corresponding data point with the function $c \cdot sinc^{-1}(\gamma \Delta G_z TE_i \, \Delta z/2)$, where $c$ is a normalisation constant and $sinc(x) = \sin(x)/x$.

**Absolute Water Content Mapping**

The standard protocol for quantitative water content mapping relies on a single exponential fit for the extrapolation of the decay curve to zero echo time (Neeb et al., 2008). However, the resulting parameter, $S_{0,T2*}$, might be biased in multiexponentially decaying tissue such as white matter. In order to evaluate possible systematic errors introduced by the single exponential model on $S_{0,T2*}$ and therefore tissue water content, Equation (1) was independently sampled 10000 times and the average and standard deviation of the extrapolated signal intensity were determined. The simulation was repeated for different slow and fast relaxation times in the interval $T_2^{*,s} \in [25ms, 85ms]$ and $T_2^{*,f} \in [10ms, 20ms]$.
In addition, the whole analysis described above for the estimation of the *optimised myelin set* was repeated. However, the goal was now to find a parameter vector which minimises the average systematic error of the total sum of amplitudes, $S_{Tot}$. The corresponding vector defines the *optimised water content set*. The accuracy and precision of $S_{0,T2*}$ and $S_{Tot}$ were determined in order to evaluate which of the two measures is more appropriate for absolute water content mapping.

**Image Processing**

Quantitative maps of $T_1$ and $T_2^*$ were reconstructed as described in (Neeb et al., 2008). However, the protocol for total water content mapping was changed in order to minimise biased results in tissue with significant biexponential decay (see also Results section). As primarily white matter exhibits biexponential behaviour, the total water content measurement for all white matter voxels was based on $S_{Tot}$ obtained from the *optimised water content set* instead of $S_{0,T2*}$, where the latter quantity is based on a single exponential model. Moreover, voxels with relative myelin water content $> 5\%$ were processed employing $S_{Tot}$. This allows for a less biased total water content mapping in other tissues with biexponential $T_2^*$ decay, such as MS lesions where the myelin sheet is still partly intact. The remaining part of the total water content reconstruction, i.e. the determination of all relevant correction factors, was left unchanged.
Furthermore, relative and absolute myelin water contents were determined for each voxel based on the linear optimisation approach described above using the *optimised myelin set*. In order to better display the relevant structures, the corresponding myelin water maps were smoothed with a Gaussian filter with kernel width of 1.5 pixels. However, all statistical analyses were based on the original, unsmoothed datasets.
For all maps, none-brain tissue was removed from all slices as follows. First, a binary map was created from the $T_1$ map where foreground pixel are defined by $T_1 > 0$. Second, tissue with $T_1 < 450ms$, $T_2^* < 25ms$ or total water content $< 60\%$ were eliminated from the binary image. Those criteria remove most of the skull visible in the MR maps without significantly effecting WM, GM or CSF. In a third step, holes were filled using a binary closing operation with a disc of radius 3 pixels as structuring element. Finally, the image was labelled based on the 4-neighbourhood relation. Pixels not belonging to the largest connected component were set to zero to remove any spurious none brain tissue which passed the above mentioned selection criteria. Furthermore, masks for white and grey matter as well as CSF were created using the $T_1$-based histogram segmentation as described in (Neeb et al, 2006b). However, the thresholds were adapted in order to account for the increased longitudinal relaxation times at 3T. Voxels with $T_1 \in [500ms, 899ms]$ were assigned to white matter while voxels with $T_1 \in [900ms, 1300ms]$ were assigned to grey matter. The CSF compartment was defined by a $T_1$ relaxation time $> 1300ms$. However, as the distribution of longitudinal relaxation times in grey matter and MS lesions is comparable, this simple approach classifies most of the lesions as grey matter. Therefore, the segmented grey matter compartment contains information from both normal appearing grey matter as well as pathological white matter lesions. Consequently, no statistical analysis was performed for the grey matter segment. For the white matter segment, average and standard deviation of total water content, absolute and relative myelin water content as well as $T_1$ and $T_2^*$ were determined separately for both patient groups investigated (EDSS=0 and EDSS=1).

The algorithms for the reconstruction and analysis of all quantitative maps as well as data simulation were implemented in MATLAB Version 7.6 (The Mathworks Inc., Natick/MA, USA) as described in (Tonkova, 2009).

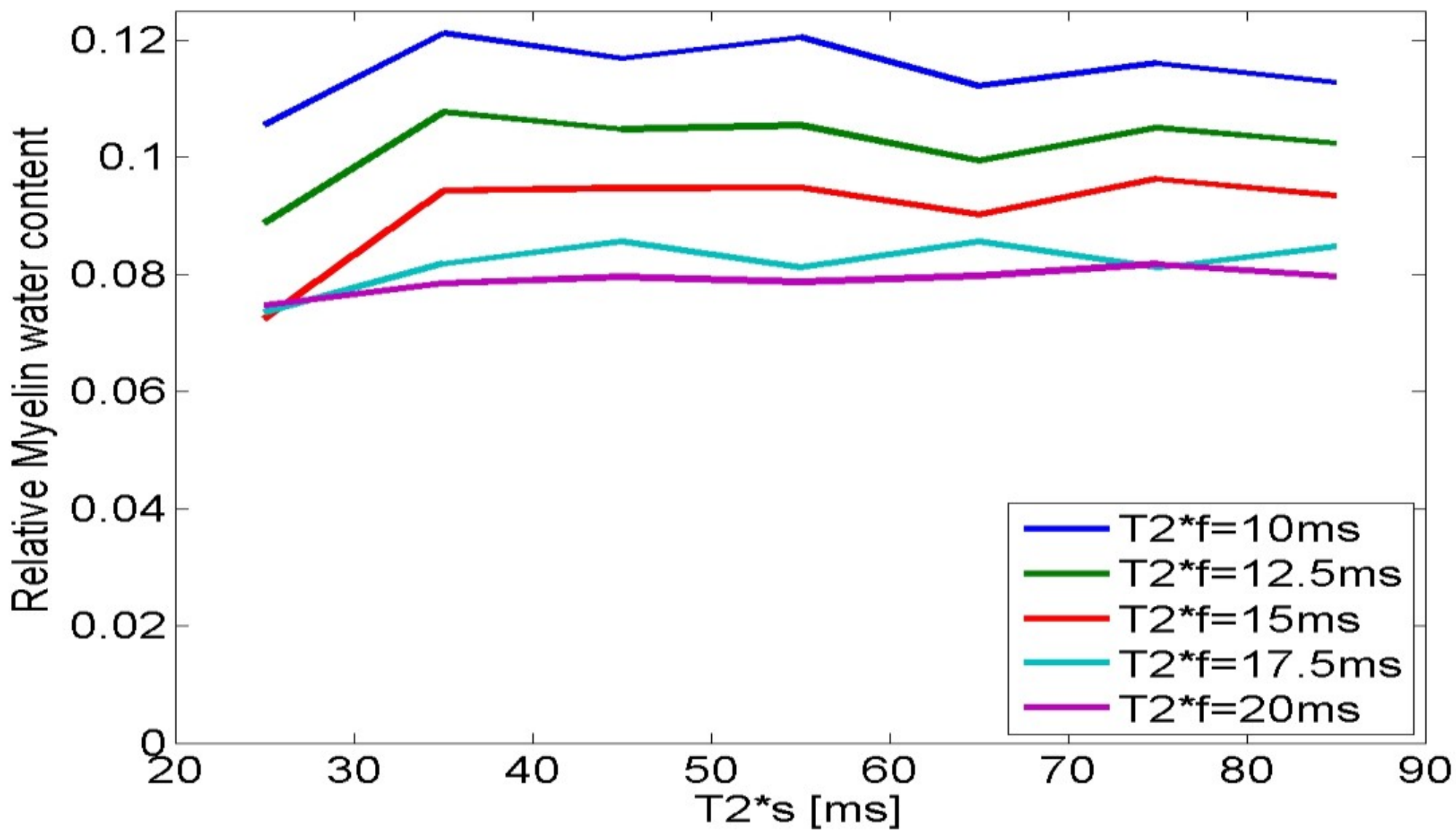

**Fig 1**: Relative myelin water content as a function of $T_2^{*,s}$ for different values of the relaxation time of the fast component, $T_2^{*,f}$.

**RESULTS**

Figure 1 shows the dependence of the relative myelin water content measurement on both fast and slow relaxation time for the *optimised myelin set*. The latter was obtained from an exhaustive search of parameter space as described above. The corresponding parameter vector, $\left(\widehat{T_2^{*,f}}, \widehat{w_f}, \widehat{w_s}\right)$, defining the *optimised myelin set*, was determined to $(10, 10, 2)ms$. Therefore, the interval width of the slow relaxation component is a factor 5 smaller than the interval width of the fast relaxation component. This is due to the fact that the position of the slower interval is adapted for each data point while $\widehat{T_2^{*,f}}$ is constant. Consequently, a larger range is required in order to constrain the systematic error also in cases where the fast relaxation time deviates from $\widehat{T_2^{*,f}}$. As can be seen from Fig.1, the myelin water content measurement is indeed much more sensitive to changes of the fast relaxation time than to variations of $T_2^{*,s}$. Changing $T_2^{*,s}$ introduces only a negligible systematic error, even for the large range considered between $25ms$ and $85ms$. On the other hand, myelin water content ranges between 0.08-0.12 for a true value of 0.1 for different fast relaxation times, $T_2^{*,f}$. Consequently, the maximum systematic error has to be estimated as $\pm 0.02$. The average absolute deviation of all simulated points from the true value was $\approx 30\%$ smaller resulting in a value of +0.0137. Furthermore, it is noticed from Fig.1 that myelin water content is overestimated for short $T_2^{*,f}$ and underestimated for long $T_2^{*,f}$.

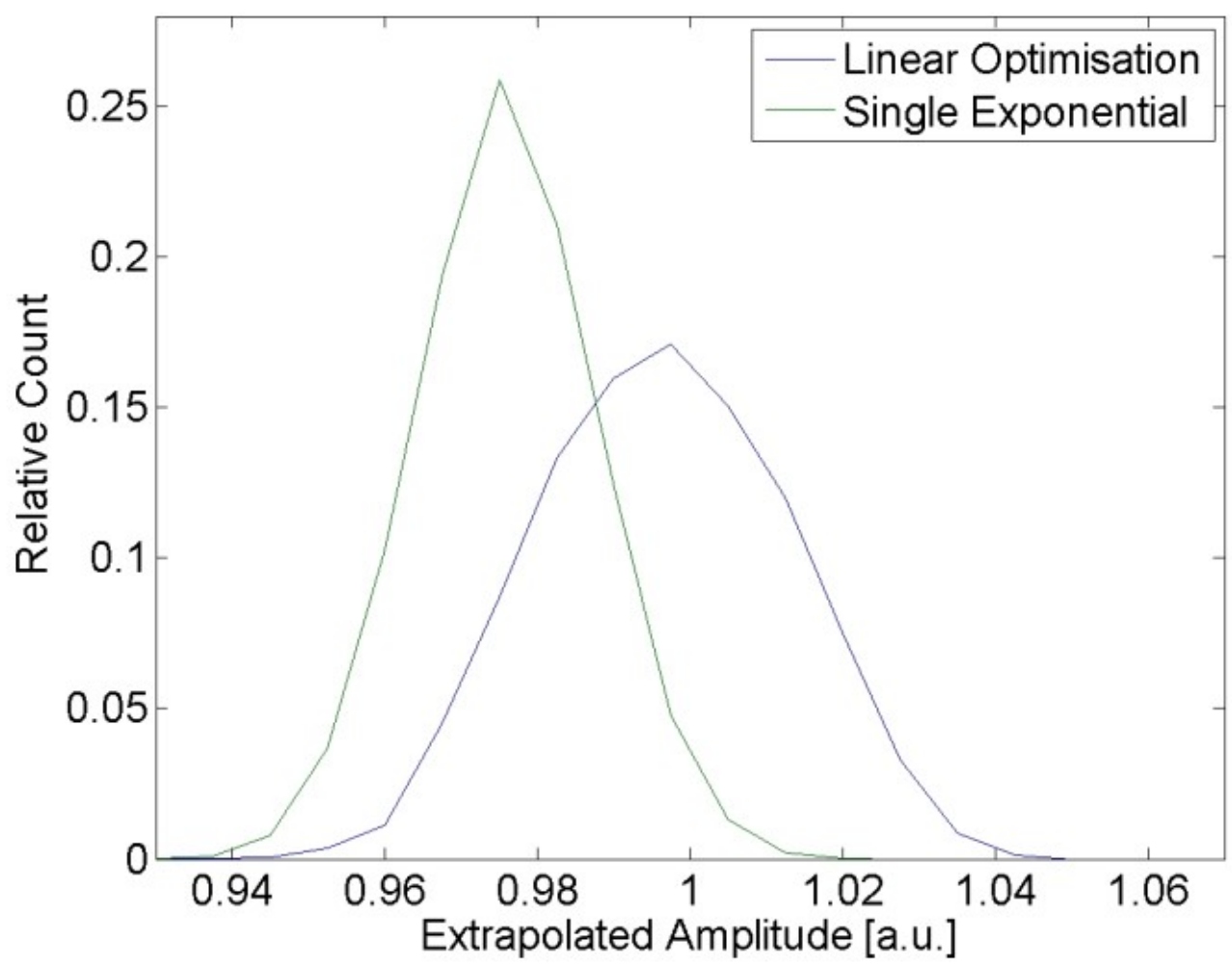

**Fig 2**: Histogram of the extrapolated signal intensity using a single exponential fit (green line) or the linear optimisation based on the *optimised water content set* as described in the text (blue line). The true signal intensity used in the simulation was set to 1.

The presence of a second, fast relaxing pool alters the determination of total water content. As can be seen from Fig. 2, extrapolation of a biexponential curve to zero echo time employing a single exponential model results in a systematic underestimation of the true value. On the other hand, the linear optimisation approach developed in the current work introduces almost no bias. The average absolute deviation of the monoexponential model was +0.021 which has to be compared to a deviation of +0.0041 observed using the LP approach. However, the statistical fluctuations are larger for the latter (0.015 vs. 0.013) as can be seen from the width of both distributions (Fig. 2). The results for the LP approach were obtained based on the *optimised water content set* with $\widehat{T_2^{*,f}} = 20ms$, $\widehat{w_f} = 10ms$ and $\widehat{w_s} = 2ms$. Therefore, only the centre position of the fast relaxation interval differs with respect to the *optimised myelin set*. Finally, the presence of a third relaxation component does not introduce a significant bias for both absolute and myelin water content measurement. The average systematic error of the myelin water content was decreased to 0.013 while the corresponding error of the total water content slightly increased to a value of 0.0044.

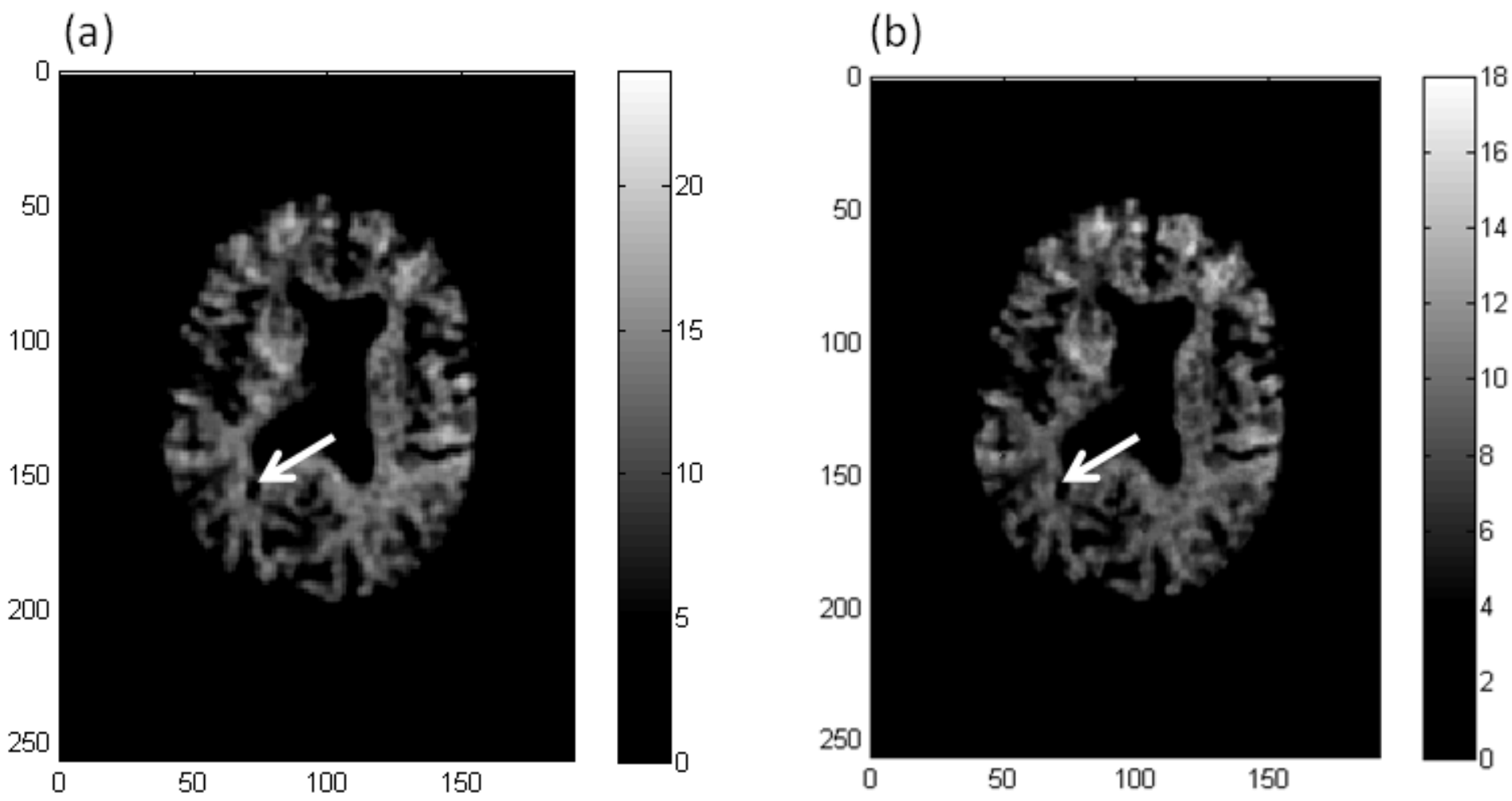

**Fig 3**: Relative (a) and absolute (b) myelin water content map for a randomly selected patient. The white arrow points to a visible MS lesion which was analysed as described in the text.

Fig. 3 shows a representative relative (a) and absolute (b) myelin water content map for a randomly selected patient. Both maps appear qualitatively comparable. I.e., it is noted on both maps that myelin water content is reduced in the periventricular lesion (white arrow). The relative myelin water content within the lesion was 4.0% in comparison to 11.2% in normal appearing white matter surrounding the lesion. The corresponding values for the absolute water content were 3.1% and 7.6%, respectively. Therefore, the reduction observed in the relative measurement is approximately 10% larger than the reduction observed in the absolute measurement resulting in an overestimation of tissue demyelination.

The average absolute white matter myelin water content in the EDSS=1 group is reduced by 2.9% with respect to the EDSS=0 group as shown in Fig. 4a and Table 1. The corresponding reduction is smaller for the relative measurement which is compatible with the slightly decreased total water content in the EDSS=1 group. Similar small changes were observed for the average $T_1$ and $T_2^*$ in white matter. However, the standard deviation of $T_2^*$ is increased by 23.9% in the group with higher EDSS score (Fig. 4b). A similar pattern is observed for both absolute and relative myelin water content where the group averaged standard deviation has increased by approx. 4%. These changes reflect stronger focal variations of the corresponding parameters whereas the average over all voxels remains to a high degree unaffected.

| | EDSS 0 | | EDSS 1 | | |
|---|---|---|---|---|---|
| | Mean | Std | Mean | Std | Relative Change |
| $T_1$ | 732.2 ms | 8.9 ms | 736.6 ms | 18.7 ms | 0.6 % |
| $T_2^*$ | 53.2 ms | 1.4 ms | 52.3 ms | 2.4 ms | -1.7 % |
| $W_{MR}$ | 70.8 % | 2.0 % | 70.6 % | 2.2 % | -0.33 % |
| $W_{My}^{rel}$ | 10.17 % | 0.57 % | 9.9 % | 0.71 % | -2.7 % |
| $W_{My}^{abs}$ | 7.16 % | 0.28 % | 6.94 % | 0.57 % | -2.9 % |
| $\sigma(T_1)$ | 86.6 ms | 4.9 ms | 87.6 ms | 9.0 ms | 1.2 % |
| $\sigma(T_2^*)$ | 9.5 ms | 1.5 ms | 11.8 ms | 1.6 ms | 23.8 % |
| $\sigma(W_{MR})$ | 4.39 % | 0.27 % | 4.33 % | 0.39 % | -1.3 % |
| $\sigma(W_{My}^{rel})$ | 3.7 % | 0.22 % | 3.85 % | 0.22 % | 4.0 % |
| $\sigma(W_{My}^{abs})$ | 2.5 % | 0.13 % | 2.61 % | 0.24 % | 4.1 % |

**Tab 1**: Mean values of $T_1$, $T_2^*$, total water content ($W_{MR}$), relative myelin water content ($W_{My}^{rel}$) and absolute myelin water content ($W_{My}^{abs}$) for the EDSS=0 and the EDSS=1 group. The corresponding standard deviations of all parameter distributions are indicated by $\sigma(...)$ and are shown in the last five rows of the table. Relative changes were calculated with respect to the value measured in the EDSS=0 group.

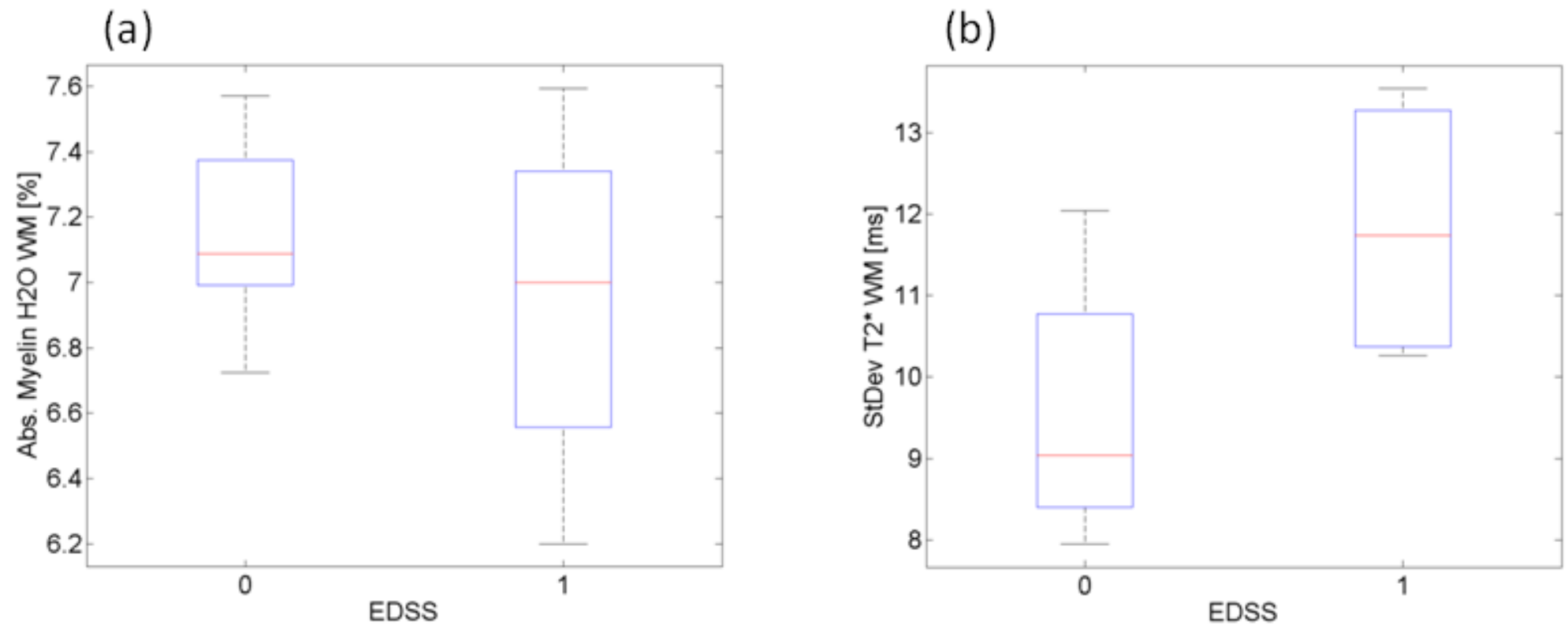

**Fig 4**: Box-Whisker Plot showing the distribution of absolute myelin water content (a) and standard deviation of the $T_2^*$ measurement in white matter (b) for the EDSS=0 and the EDSS=1 group.

## DISCUSSION AND CONCLUSION

Multiple sclerosis is one of the diseases were the application of quantitative MRI in diagnosis and monitoring is very actively discussed by many researchers. This is basically due to two characteristics of quantitative MRI which makes its use very attractive for research and diagnosis of MS. First, multiple sclerosis is a long term disease which is typically diagnosed in the first decades of live and where currently no causal treatment options are available. Therefore, patients are undergoing regular clinical MR scans in order to monitor the disease progression. However, environmental conditions such as scanner hard- and software undergo their natural product cycle making an objective comparison of imaging results between different timepoints difficult. Quantitative MRI on the other hand might allow for a better and more objective comparison as the results obtained depend on physical and physiological properties of each subject only. Furthermore, it is expected that a more direct and quantitative assessment of parameters which display pathological processes might help to resolve the observed mismatch between clinical and imaging appearance of patients (Bakshi et al., 2008). The latter aspect is especially important for the monitoring of therapeutic response which heavily relies on the availability of objective information relevant for the prognosis of further disease progression. Consequently, many different approaches to measure e.g. diffusion anisotropy, relaxation properties or myelin water content have been investigated in the literature for an improved assessment of multiple sclerosis (see e.g. Bakshi et al. 2008; Neema et al., 2007 for an overview). However, those methods often suffer from one or more of the following drawbacks, which might possibly hinder a more widespread use of quantitative MRI: (1) limited number of parameters mapped simultaneously, (2) limited resolution or spatial coverage, (3) dedicated sequences which are available only in a few institutions and/or (4) long acquisition time.

In order to overcome those problems, we have extended the previously reported method for simultaneous mapping of $T_1$, $T_2^*$ and absolute water content with whole brain coverage (Neeb et al., 2008). The current work was primarily focused on feasibility of myelin water content mapping using the same protocol without necessity to acquire additional data. To optimize the information obtained per unit measurement time, the best compromise between spatial coverage, measurement time and a most complete sampling of the $T_2^*$ decay curve has to be found. Unfortunately, however, fulfilling all requirements simultaneously would lead to none overlapping regions in sequence parameter space and therefore to contradictory measurement protocols. In the current work, spatial coverage and $T_{acq}$ were optimised for a protocol used in a different context, so that the sampling density was naturally compromised to only 10 points on the relaxation curve with a limited $TE$ range ($TE_{max} < 40ms$). Therefore, our approach is expected work especially well in gradient echo experiments due to the faster signal decay as compared to spin echo acquisitions. However, gradient echo sequences are sensitive to static magnetic field inhomogeneities which are primarily observed close to tissue air interfaces. Here, the separation between fast and slow component might become too small to reliably resolve both peaks. This typically results in an overestimation of the fast component and therefore myelin water content. Moreover, the decay signal deviates from a pure single or multi exponential behaviour if large magnetic field gradients are present (Neeb et al., 2006a). In such regions, the decay model applied is no longer valid rendering the estimation of pool-specific amplitudes difficult. Therefore, we have corrected our decay data prior to the optimisation process by assuming a linear variation of the magnetic field strength across a voxel with constant proton density. One has to be careful however, as this assumption might introduce a systematic error if the field does not vary linear or if the proton density is heterogeneously distributed. Nevertheless, as previously shown, only a small part of all voxels in the brain require a correction for such off-resonance effects (Neeb et al., 2008), so that the overall systematic error introduced in such regions by a possibly improper correction might be neglected, especially if one is primarily interested in an averaged behaviour.

Even though $T_2^*$ decay proceeds faster than $T_2$ decay, especially at higher field strength, the sparse sampling of the relaxation curve and the more densely spaced peaks significantly increase demands on the proper choice of the best optimisation approach. In the current work, we were dealing with this problem by employing linear programming (LP) optimisation. LP offers much more flexibility to include prior information or assumptions about the relaxation process, which is not feasible using nonlinear least squares optimisation. Based on the two pool model already successfully applied by many researchers (Laule et al., 2004; Tozer et al., 2005; Laule et al., 2006; Oh et al., 2007), this flexibility was employed to define two distinct relaxation intervals where the optimisation was performed. This naturally raises the question about systematic errors introduced by the specific choice of relaxation intervals. We have tried to minimise any possible bias by using as much information as possible contained in the measured data in order to define and constrain the six parameters defining maximum amplitude, position and width of both relaxation intervals. Due to the dominating effect of the long relaxation compartment on the observed relaxation curve, simulations have shown that $\widehat{T_2^{*,s}}$ can well be approximated using information gathered from a single exponential fit. Furthermore, constraints of the maximum compartment amplitudes were extremely weak. We have required that the fast relaxation pool amplitude should not exceed 50% of the total amplitude. Given the typical distribution of myelin water content in the human brain previously reported (0-20%), any systematic bias introduced by this choice can almost certainly be neglected. The remaining three parameter were systematically varied in order to minimise the average systematic error over a large range of $T_2^{*,f}$ and $T_2^{*,s}$. As demonstrated, only the fast relaxation time has a significant influence on the observed systematic error. Unfortunately, not very much is known about the real distribution of $T_2^{*,f}$ in cerebral tissue. Thus, a measured myelin water content of 0.1 has to be conservatively estimated with a maximum error bar of $\pm 0.02$ for a single voxel. Nevertheless, the deviation from the true value, averaged over the whole range of $T_2^{*,f}$ investigated was smaller (0.013). Any prior information to constrain the range of $T_2^{*,f}$ will obviously help to further reduce the systematic error. Such information might be obtained either from additional measurements with a higher sampling density of the relaxation curve or using data driven methods in order to estimate $\widehat{T_2^{*,s}}$. In contrast to the dependence on the fast relaxation time, the systematic error is almost independent of $T_2^{*,s}$. This observation again stresses the fact that the heuristic approach employed here to define the position of the slow relaxation interval works extremely well.

Furthermore, the investigation of systematic shifts introduced by a potential third relaxation component with long $T_2$ has shown that both myelin and total water content are almost left unaffected. A long $T_2$ relaxation component has recently been observed in normal appearing white matter of patients suffering from phenylketinuria and multiple sclerosis as well as healthy controls (Laule et al., 2007a). The observation of a constant average systematic error does not necessarily imply that the myelin water content measurement is immune against the presence of a third relaxation component. However, we can conclude that other sources, i.e. the unknown fast relaxation time, dominate the total systematic error so that effects from a long relaxation compartment can be neglected.

The above given conservative estimates for the systematic error refer to the measurement of myelin water content within a single voxel. However, if one is primarily interested in the average myelin water content of white or grey matter, the estimate has to be refined. In this case, the maximum systematic error given above would be observed only in subjects with a constant, voxel independent fast relaxation time, $T_2^{*,f}$. However, such an assumption seems to be quite unlikely. It would require e.g. constant diffusion properties across white matter as the diffusion coefficient is correlated with the transverse relaxation time. Furthermore, field inhomogeneities result in a modulation of the relaxation curve in a gradient echo experiment even in regions with comparable microstructural composition. Therefore, it is not very likely that the maximum systematic error observed in the simulation study is a proper estimate for the real systematic error of the average myelin water content *in vivo*. In order to obtain a more realistic estimate of the bias, one needs to make assumptions about the true distribution of the fast relaxation time. It is important to notice that the simulation study has shown a symmetric deviation from the true myelin water content with respect to $T_2^{*,f}$ changes. To be more concrete, myelin water content will be overestimated with shorter $T_2^{*,f}$ and underestimated with longer $T_2^{*,f}$. Assuming e.g. a flat distribution of $T_2^{*,f}$, the systematic error would be close to zero as negative deviations are almost perfectly compensated by positive deviations. However, as not much is known about the real distribution of short $T_2^*$ in the human brain, one should refer to measured data as much as possible. Here, it is first noticed that the average $T_2^*$ measured in the current study is decreased by 1.7% in the EDSS=1 group with respect to the group with EDSS=0. In an extreme case, one could assume that this effect is purely due to a smaller $T_2^{*,f}$ with constant $T_2^{*,s}$. However, it might be more likely that changes in both pools contribute to the reduced average $T_2^*$ in the EDSS=1 group, e.g. due to field inhomogeneities which affect both fast and slow relaxing components in a similar way. Still, the observed $T_2^*$ differences cannot explain the observed difference in myelin water content. Decreasing e.g. $T_2^{*,f}$ results in an overestimation of myelin water content as can be seen from Fig. 1. Nevertheless, myelin water content was measured to be reduced and not increased so that the actual reduction observed for the EDSS=1 group might even be slightly larger.

The determination of total water content based on multi-echo datasets requires a proper model to extrapolate the decay curve to zero echo time. The presence of a pool with fast relaxation time of $\sim 10 - 15 ms$ results in an underestimation of water content by $\sim 1.5 - 2$ percent points even for a small pool fraction of 10%. Therefore, previously published results based on multi-echo measurements should be reviewed critically. Even though grey matter measurements might be not have been significantly biased, total white matter water content is very likely to be underestimated in cases where a single exponential model has been employed (Lin et al., 2000; Venkatesan et al., 2000; Warntjes el al., 2007; Warntjes el al., 2008; Neeb et al., 2008). For this reason, the determination of total tissue water content was modified in the current study. For white matter, it is *a priori* assumed that a fast relaxation pool due to myelin bound water is present so that the signal intensity at $TE = 0$ is determined by $S_{Tot}$, the sum of amplitudes of all relaxation pools present. The corresponding increase in accuracy is partly compromised by a decreased precision due to the slightly lower SNR of $S_{Tot}$ with respect to the result obtained by a single exponential fit. Therefore, the multi pool analysis was restricted to voxels where it was necessary to employ this model to obtain unbiased results. I.e., this refers to voxels where a significant biexponential behavior was either assumed (such as white matter) or has been observed, such as MS lesions without significant demyelination.

Multidimensional quantitative MRI (mqMRI) in MS might especially be suited for the study of (1) longitudinal changes in an individual patient during the natural course of disease and (2) for a sensitive monitoring of changes induced by therapeutic interventions. For this reason, we have restricted our cohort to volunteers already diagnosed to suffer from MS in order to demonstrate the feasibility to detect changes in myelin and total water content as well as $T_1$ and $T_2^*$ of the human brain using the approach developed in the current work. We have observed a small reduction of absolute myelin water content by 2.9% in the EDSS=1 group with respect to the EDSS=0 group. However, the overlap between both groups is relatively strong so that statistical significance is limited given the small sample size of the current study. Nevertheless, a reliable measure of myelin water content can be obtained using the approach developed as the results ($W_{My}^{rel} \approx 10\%$) are consistent with results obtained by others (Laule et al., 2004; Oh et al., 2007; Du et al., 2007; Sirrs et al., 2007; Kolind et al., 2009; Vavasour et al., 2009; Hwang et al., 2010). In addition to the relative measure of myelin water content, an absolute quantification was performed which allows for a more precise measure in the presence of total water content changes. This is primarily relevant for the assessment of myelination in MS lesions. As most lesions are associated with increased total water content, presumably of the free pool due to oedema, a relative measurement of myelin water content overestimates the corresponding demyelination process. We have shown in randomly selected patient that indeed the reduction of absolute myelin water content relative to surrounding normal appearing matter was approximately 10% smaller than the result obtained for the relative measurement. This is exactly the rise in total water content typically observed in a lesion. Even for lesions of a single subject, however, we have observed a considerable spread of water content increase, ranging from 0% up to 25%. Therefore, an accurate measure of lesion myelin water content for a comparison between different lesions of a single subject or for the longitudinal follow up of a single lesion should be based on an absolute measurement of myelin water content. The same argument applies for the assessment of normal appearing white matter, although to a lesser extent. Even though average changes of white matter water content are much smaller than the corresponding changes observed in lesions, regional variations may be considerable. As previously demonstrated in a study of healthy individuals, the regional distribution of cerebral total water content significantly depends on age and gender (Neeb et al., 2006b). Therefore, similar effects might be observed in diseased subjects. Moreover, myelin water content mapping is especially attractive for the depiction of early demyelination processes not yet visible on conventional MRI. Here, is especially important to account for the cofounding effect of accompanied oedema due to inflammation as changes in myelin water content are typically small. A small overestimation of a small change results in a large systematic error which can significantly be reduced by a simultaneous measurement of total water content.

To our knowledge, this is the first study where total water content maps for the whole brain were acquired in patients suffering from multiple sclerosis. The results obtained are consistent with previous studies in a healthy population (Neeb et al., 2006b; Neeb et al., 2008), indicating that significant global oedema is not present in MS patients with moderate disease. Furthermore, total water content, $T_1$ and $T_2^*$ did not vary significantly between the EDSS=0 and EDSS=1 group. This is consistent with previous reports, where only small to moderate correlations between $T_1$ in white matter and clinical score were described (Vaihianather et al., 2002, Parry et al., 2002). Given the fact that pathological differences between two groups with similar clinical appearance might be relatively small, a global analysis requires a much larger sample size in order to significantly detect changes. Nevertheless, the average white matter $T_1$ found in the current study is consistent with other measurements performed at 3T (Vrenken et al, 2006a). The missing correlation between EDSS and mean $T_1$, $T_2^*$ and total water content does not exclude regional variation of those parameters. A simple and reproducible measure of increased regional variations is provided by the standard deviation of each quantitative parameter. Here, it is especially noted that the standard deviation of $T_2^*$ is increased by almost 24% in the EDSS=1 group, indicating a larger spatial heterogeneity of $T_2^*$ without relevant change of the average value. This result is statistically significant even for the small population studied in the current work. Vrenken et al. (2006a) have reported about similar changes in the width of the $T_1$ histogram between healthy controls and MS subjects, where patients with secondary progressive MS showing the largest increase in peak width. An inverse relationship between disease duration and $T_1$ peak height in normal appearing white matter was found in (Parry et al., 2003). Furthermore, significant age and gender dependent regional variations in absolute water content maps were observed and quantified based on characteristic image features in a healthy control group (Neeb et al., 2006b). Therefore, the observed increase of the $T_2^*$ standard deviation might present a similar physiological effect with an increased heterogeneity in normal appearing white matter. This might, at least in parts, be explained by the observed increase in the standard deviation of the absolute myelin water content measurement from EDSS=0 to EDSS=1, reflecting a less homogeneous myelination throughout white matter in the latter group. However, some care must be taken when interpreting $T_2^*$ data as different shimming conditions or local susceptibility gradients might account for some variation independent of real physiological effects. As described above, our data were corrected for magnetic field variations so that changes due to off resonance sources can be neglected to 1$^{st}$ order. Nevertheless, it cannot be excluded that part of the observed increases of the $T_2^*$ and myelin water content standard deviations might be attributed to remaining higher order terms.

The simultaneous acquisition of multiple parameters using a single protocol naturally leads to the application of multivariate analysis strategies. One of the key advantages over a sequential univariate analysis is the possibility to include higher dimensional correlations between parameters. Therefore, a multivariate disease classifier is based on additional information which typically increases its performance over its univariate counterparts. It can thus be expected, that e.g. an improved prediction of EDSS score might be attainable based on a multidimensional combination of $T_1$, $T_2^*$ as well as total and myelin water content. However, one important aspect of data reconstruction hinders the naïve application of such approaches. As all four parameters measured are based on the same datasets, artificial correlations between parameters occur. This is due to the propagation of noise from the original data to the reconstructed maps. In order to demonstrate this effect in a simple model, we assume that the first point acquired on the relaxation curve fluctuates above its real value while all other points are measured without error, i.e. they coincide with their real values. In this situation, fitting the relaxation data would result in an overestimation of the extrapolated signal intensity at $TE = 0$ and therefore tissue water content. At the same time, however, $T_2^*$ decay appears to proceed faster, resulting in an apparently negative correlation between $T_2^*$ and total water content. The issue of artificial data correlations is currently investigated at our institution in order to allow for the application of multivariate analysis techniques of quantitative MR parameters in multiple sclerosis.

In conclusion, we have demonstrated that tissue $T_1$, $T_2^*$ as well as total and myelin water content can be mapped simultaneously with whole brain coverage in less than 10 minutes and $1x1x2mm^3$ voxel size. In contrast to previous reports on myelin water content mapping, our approach allows for an absolute assessment of myelin water content. Therefore, cofounding effects from the simultaneous presence of oedema are eliminated. Further speed gain can be achieved using parallel acquisition techniques. Without specific requirements on hardware or measurement sequence, the approach can be applied on almost every clinical scanner. This might help to expedite the further spread of quantitative MRI for both research and diagnosis of multiple sclerosis.